\documentclass{webofc}

\usepackage[varg]{txfonts}   
\usepackage{hyperref}
\usepackage{url}

\usepackage{lineno}
\usepackage{multicol}
\usepackage{amsfonts}
\usepackage{amsmath}
\usepackage{array}
\usepackage{amssymb}
\usepackage{mathrsfs}
\usepackage{graphicx}
\usepackage{dcolumn}
\usepackage{longtable}
\usepackage{multirow}
\usepackage{epstopdf}
\usepackage{natbib}
\usepackage{subfigure}
\usepackage{indentfirst}
\usepackage{diagbox}
\usepackage{float}
\usepackage{mathtools}
\usepackage{picinpar}
\hypersetup{colorlinks=true,citecolor=blue,urlcolor=blue,linkcolor=blue}
\begin{document}
%

\title{Directed Flow of $\Lambda$, $\prescript{3}{\Lambda}{\rm H}$ and $\prescript{4}{\Lambda}{\rm H}$ in Au+Au collisions at $\sqrt{s_{NN}}$ = 3.2, 3.5, 3.9 and 4.5 GeV at RHIC}
%
%

\author{\firstname{Junyi} \lastname{Han}\inst{1,2}\fnsep\thanks{\email{JYHan@mails.ccnu.edu.cn}} (for the STAR collaboration)
}

\institute{Central China Normal University
\and
           Heidelberg University
          }

\abstract{
Studying hyper-nuclei yields and their collectivity can shed light on their production mechanism as well as the hyperon-nucleon interactions. Heavy-ion collisions from the RHIC beam energy scan phase II (BES-II) provide an unique opportunity to understand these at high baryon densities. \\
\indent In these proceedings, we present a systematic study on energy dependence of the directed flow ($v_{1}$) for $\Lambda$ and hyper-nuclei ($^{3}_{\Lambda}{\rm H}$, $^{4}_{\Lambda}{\rm H}$) from mid-central Au+Au collisions at $\sqrt{s_{NN}}$ = 3.2, 3.5, 3.9 and 4.5 GeV, collected by the STAR experiment with the fixed-target mode during BES-II. The rapidity (y) dependence of the hyper-nuclei $v_{1}$ is studied in mid-central collisions. The extracted $v_{1}$ slopes ($dv_{1}/dy|_{y=0}$) of the hyper-nuclei are positive and decrease gradually as the collision energy increases. These hyper-nuclei results are compared to that of light-nuclei including p, d, t/$\rm ^{3}He$ and $\rm ^{4}He$. Finally, these results are compared with a hadronic transport model including coalescence after-burner.
}
\maketitle
%

\section{Introduction}
\label{intro}
\indent Heavy-ion collisions (HICs) are a powerful tool to explore the properties of strongly interacting QCD matter. An extremely high temperature and high density matter is produced during HICs. 
The interations between the constituents produced in HICs will lead to the formation of bound states.
Hypernuclei are bound states of nucleons and hyperons. In astrophysics, the "hyperon puzzle" shows it is difficult to reconcile the measured masses of neutron stars with the presence of hyperons in their interiors. 
Hypernuclei provide a good opprotunity to study the hyperon-nucleon(Y-N) interactions in the QCD equation of state at high baryon density. Understanding hyperon-nucleon(Y-N) interaction in high density region is essential for solving the hyperon puzzle. 
The production mechanism of hypernuclei is also a topic of importance to understand. \\
\indent From the thermal model calculations of light-nuclei and hyper-nuclei, their production are enhanced at high baryon density region \cite{Ref_model1, Ref_model2}. 
In the second phase of the RHIC Beam Energy Scan program \cite{Ref_qcd, Ref_qcd1}, the collision energy for fixed-target Au+Au collisions is extended down to $\sqrt{s_{NN}}$ = 3 GeV. 
It gives us a good opportunity to study nucleon-nucleon(N-N) and hyperon-nucleon(Y-N) interactions by analyzing light-nuclei and hyper-nuclei.
Collective flow of hyper-nuclei can help us to explore its production mechanism and hyperon interactions in the medium. \\
\indent 

\section{Datasets and Particle Reconstruction}
\label{sec-1}
\indent 
The datasets used in this analysis were collected by the upgraded STAR experiment at RHIC with the fixed-target mode setup at $\sqrt{s_{NN}}$ = 3.2, 3.5, 3.9 and 4.5 GeV.
STAR has upgraded the inner Time Projection Chamber (iTPC), endcap Time-of-Flight (eTOF) and Event Plane Detector (EPD) \cite{Ref_itpc_etof_epd} in recent years. These upgrades bring larger acceptance, better particle identification ability with uniform efficiency and higher event plane resolution. \\
\indent 
The reaction plane is estimated via the reconstruction of the event plane in the EPD \cite{Ref_ep}. 
Three sub-event method is used to determined the EPD event plane resolution.
In 5-40\% centrality, we can both have high hyper-nuclei yield and high event plane resolution. The event plane resolution can reach over 70\% at 3.2 GeV. \\
\indent In order to reconstruct hyper-nuclei, we need the $\rm{\pi^{-}, p, d, \prescript{3}{}{He} \ and \prescript{4}{}{He}}$. These charged particles can be selected based on the ionization energy loss (dE/dx) measured in the TPC \cite{Ref_tpc} as a function of rigidity. For the short-lived particle like hyper-nuclei, they are reconstructed with KFParticle package based on Kalman filter method to improve signal significance \cite{Ref_kfp}. 
The following decay channels are used: $\rm{\Lambda \rightarrow p + \pi^{-}}$, $\rm{\prescript{4}{\Lambda}{H} \rightarrow \prescript{4}{}{He} + \pi^{-}}$, $\rm{\prescript{3}{\Lambda}{H} \rightarrow \prescript{3}{}{He} + \pi^{-}}$ and $\rm{\prescript{3}{\Lambda}{H} \rightarrow p + d + \pi^{-}}$. 
Combinatorial backgrounds for hyper-nuclei is estimated by rotating one of the daughter tracks by random angle in transverse plane. 


\begin{figure}[h]
    \centering
    \includegraphics[width=0.55\linewidth]{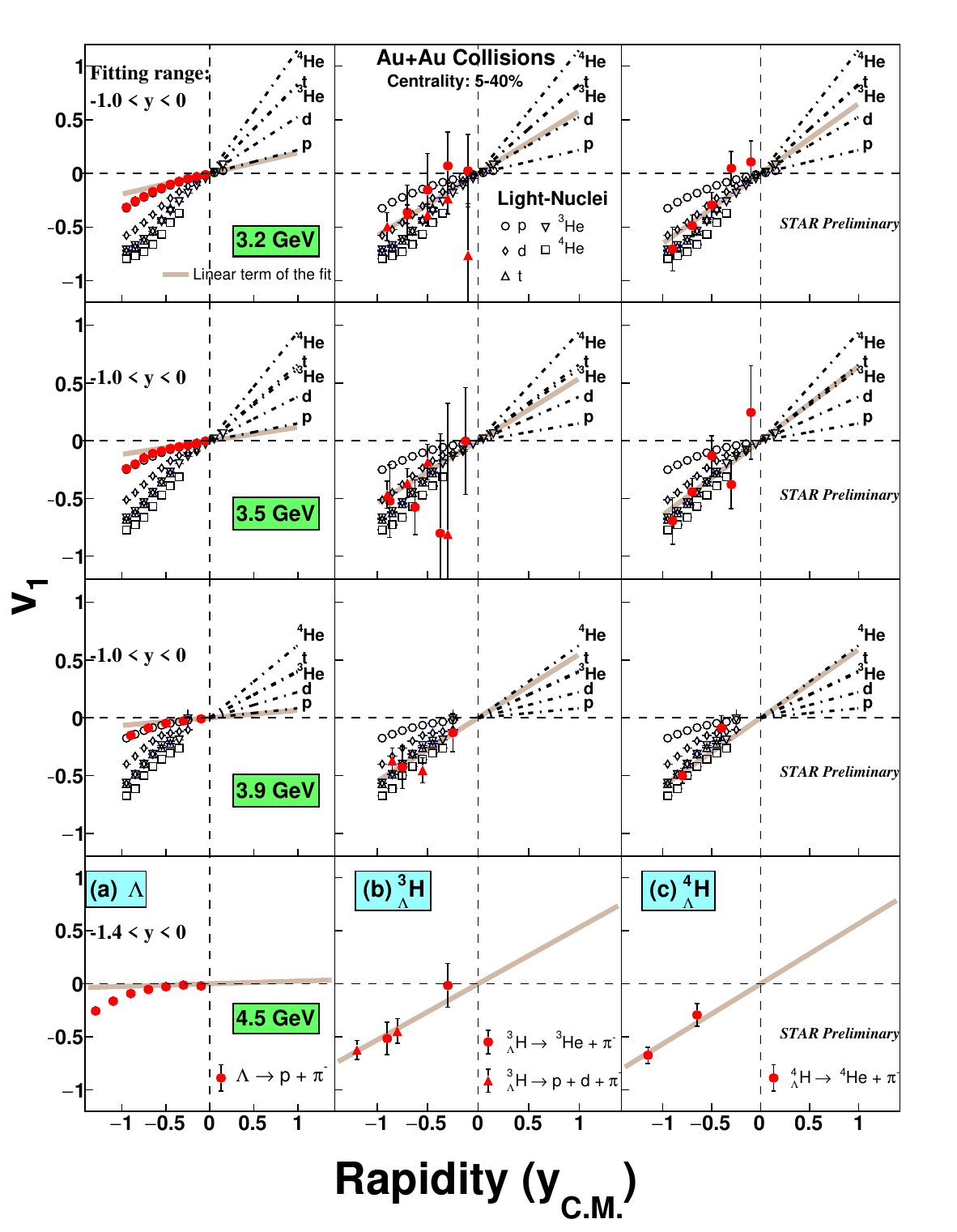}
    \vspace{-1mm}
    \caption{Light-nuclei and hyper-nuclei directed flow $v_{1}$ as a function of center-of-mass rapidity at 3.2, 3.5, 3.9 and 4.5 GeV with 5-40\% centrality Au+Au collisions \cite{Ref_v1_ep}. The rapidity dependence of $v_{1}$ for $\Lambda$ on the left, $\rm{\prescript{3}{\Lambda}{H}}$ 2-body and 3-body decay in the middel, $\rm{\prescript{4}{\Lambda}{H}}$ results on the right represented by solid markers. The linear terms of the fitting for $\Lambda$, $\rm{\prescript{3}{\Lambda}{H}}$ and $\rm{\prescript{4}{\Lambda}{H}}$ are represented as brown lines. For comparison, the rapidity dependence of $v_{1}$ for p, d, t, $\rm{\prescript{3}{}{He}}$ and $\rm{\prescript{4}{}{He}}$ are also shown as open markers, and the linear terms of the fitting results are represented as dashed lines in the positive rapidity region.}
    \label{fig:v1_y}
    \vspace{-6mm}
\end{figure}

\indent Systematic uncertainties of the measured collective flow mainly come from the event plane resolution, reconstruction efficiency, particle identification cuts and topological variable cuts. 
In case of hyper-nuclei, the dominant source of systematic uncertainty arises from the topological cuts (used for their selection).

\section{Results and Discussion}
\label{sec-2}

\indent The $v_{1}$(y) of $\Lambda$ hyperon and hyper-nuclei are extracted with event plane method \cite{Ref_v1_ep}. 
The distributions of light-nuclei and hyper-nuclei $v_{1}(y)$ from 5-40\% mid-central Au+Au collisions at 3.2, 3.5, 3.9 and 4.5 GeV are shown in the Fig.\ref{fig:v1_y}. 
Since collective flow depends on $p_{T}$, we select a similar $p_{T}/A$ range (approximately 0.4 < $p_{T}/A$ < 0.8 GeV/c) for comparison for a given particle.
As we can see, the $\Lambda$ hyperon $v_{1}(y)$ distribution is close to that of the proton, and hyper-nuclei $v_{1}(y)$ distributions are also close to those light-nuclei with the same mass numbers. 
For the $\Lambda$ hyperon and light-nuclei, we use the third-order polynomial $v_{1}(y)$ = $p_{0}y + p_{1}y^{3}$ to extract the $v_{1}$ slope ($p_{0}$ = $(v_{1})_{slope}$ = $dv_{1}/dy|_{y=0}$). 
Due to poor statistics for $\rm{\prescript{3}{\Lambda}{H}}$ and $\rm{\prescript{4}{\Lambda}{H}}$, we use the first order polynomial $v_{1}$(y) = $p_{0}y$ to fit, the fitting functions are shown by the solid brown lines. 

\begin{figure}[h]
    \centering
    \vspace{-2mm}
    \includegraphics[scale=0.45]{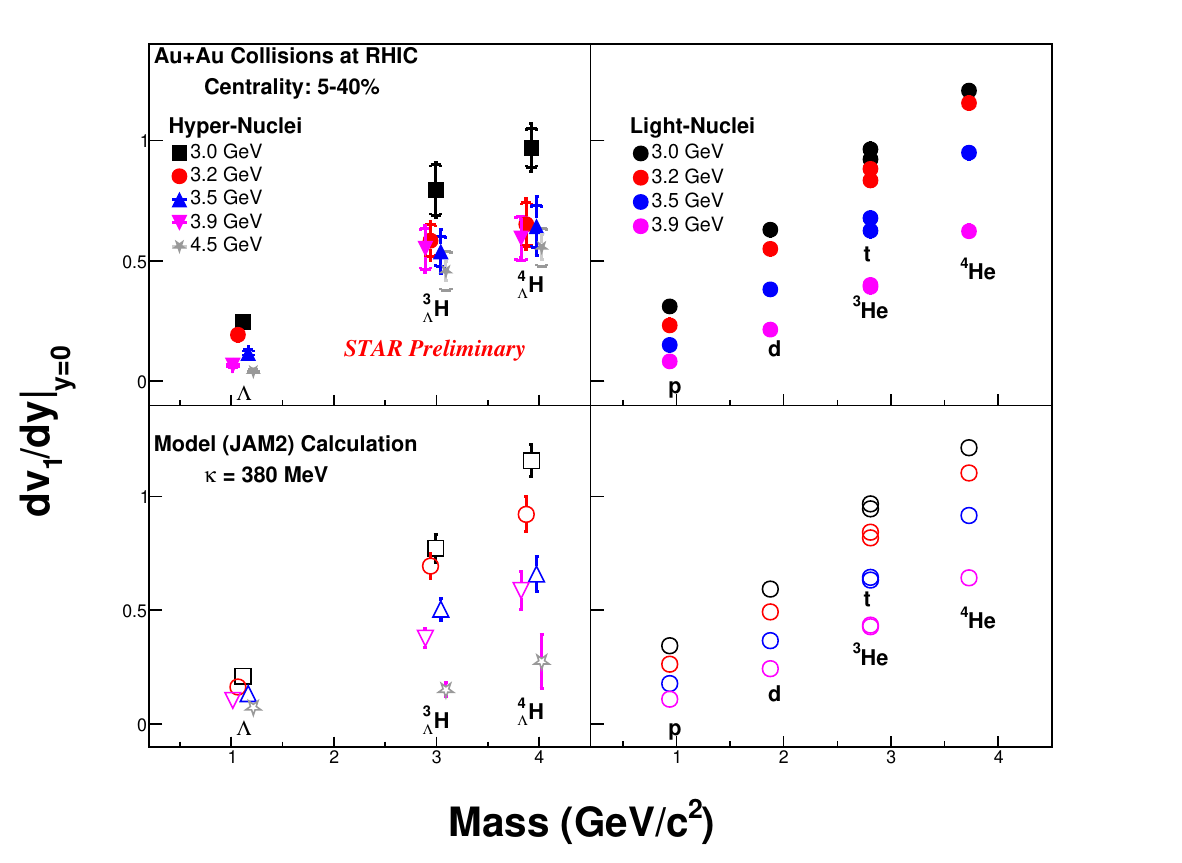}
    \vspace{-1mm}
    \caption{Particle mass dependence of the mid-rapidity $v_{1}$ slope for light-nuclei and hyper-nuclei at 3.2, 3.5, 3.9 and 4.5 GeV with 5-40\% centrality Au+Au collisions. The top results are from experimental data for hyper-nuclei(left) and light-nuclei(right), and the bottom results with open markers are corresponding results from hadronic transport model (JAM2) with coalescence afterburner calculation.}
    \label{fig:mass_dependence}
    \vspace{-3mm}
\end{figure}

\indent The results of the mid-rapidity $v_{1}$ slope as a function of particle mass for light-nuclei and hyper-nuclei with 5-40\% centrality are shown in Fig.\ref{fig:mass_dependence}. 
The top panels present results from hyper-nuclei and light-nuclei, and the bottom panels with the open markers present the hadronic transport model (JAM2) \cite{Ref_jam} with coalescence afterburner calculation results. 
As we can see, at given energy, for both light-nuclei and hyper-nuclei, the mid-rapidity $v_{1}$ slopes are scaled with particle mass, implying that coalescence is the major process for the light-nuclei and hyper-nuclei production. And the feature is also reproduced by JAM2 model calculations. 

\begin{figure}[h]
    \centering
    \includegraphics[scale=0.50]{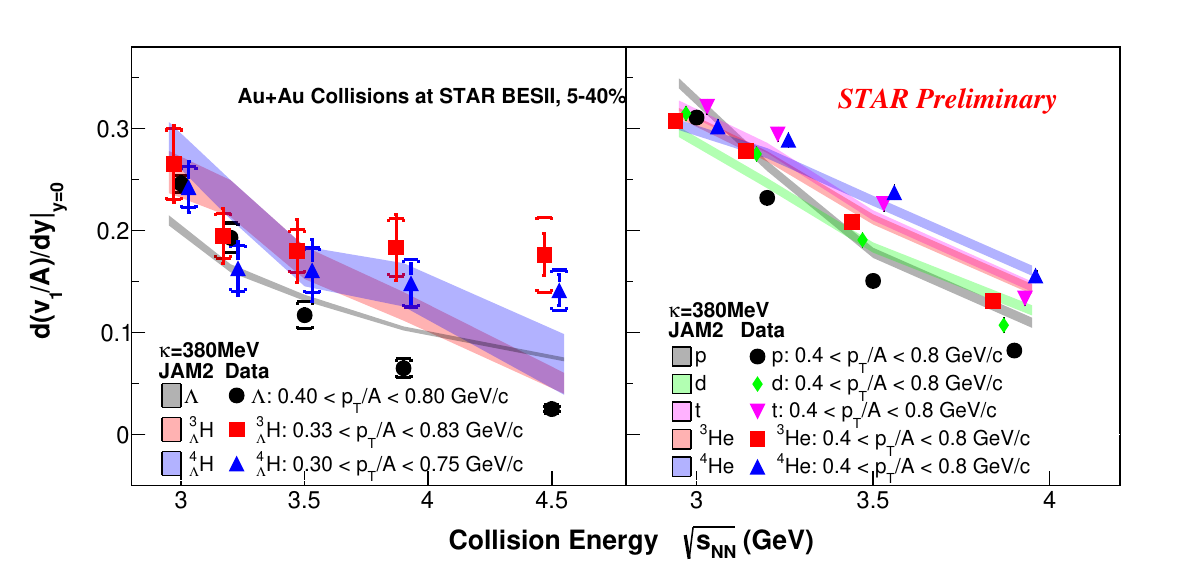}
    \caption{Collision energy dependence of the mid-rapidity $v_{1}$ slope for light-nuclei and hyper-nuclei at 3.2, 3.5, 3.9 and 4.5 GeV with 5-40\% centrality Au+Au collisions. The left results are from hyper-nuclei and the right are light-nuclei. The JAM2 model calculation results are represented by colored bands both for light-nuclei and hyper-nuclei.}
    \label{fig:energy_dependence}
    \vspace{-5mm}
\end{figure}

\indent The collision energy dependence of the mass scaled mid-rapidity $v_{1}$ slope for light-nuclei and hyper-nuclei with 5-40\% are shown in Fig.\ref{fig:energy_dependence}. 
From the results, we observe that as the collision energy increases, the mid-rapidity $v_{1}$ slope of light-nuclei and hyper-nuclei have a downward trend.
For comparison, hadronic transport model calculations with a coalescence afterburner are also included and are found consistent with the observed energy dependence in the data.

\section{Summary and conclusions}
\label{sec-3}
\indent In this analysis, we presented the particle mass and collision energy dependence of the directed flow $v_{1}$ for light-nuclei and hyper-nuclei in $\sqrt{s_{NN}}$ = 3.2, 3.5, 3.9 and 4.5 GeV Au+Au collisions at RHIC-STAR. 
The distribution of particle mass and collision energy dependence of the mass scaled mid-rapidity $v_{1}$ slope of light- and hyper-nuclei shows that as the collision energy increases, the baryon number scale breaks down, indicating that the coalescence process dominates the production mechanism of the light clusters becoming weaker.
Additionally, the hadronic transport model calculations are consistent with the observed mass and energy dependence.
STAR has collected 2 billion events from 3.0 GeV Au+Au collisions. Analysis of this data will provide improved precision, leading to stronger constraints on the coalescence parameters for both light-nuclei and hyper-nuclei.
\newline
\begin{acknowledgement}
\textbf{Acknowledgement}: This work was supported in part by the National Natural Science Foundation of China (Grant No. 12375134), the National Key Research and Development Program of China under Grant No. 2020YFE0202002, and the Fundamental Research Funds for the Central Universities (Grant No. CCNU22QN005).
\end{acknowledgement}

\end{document}